# Self-organization behavior in a constrained minority game

Ching Liu[1] and Sy-Sang Liaw[2]

[1] General Education Center, China Medical University, 91 Hsueh-Shih Road, Taichung, Taiwan

[2] Department of Physics, National Chung-Hsing University, 250 Guo-Kuang Road, Taichung, Taiwan

## Abstract

In the standard minority game, every agent switches to his best strategy in hand at each time step. If only a small number of agents are allowed to switch their strategies at each time step, the population variance of the system plunges. The variance reaches a low value and remains steady for a period of time. Then without any sign it starts to rise abruptly to a high value and then decreases smoothly to the previous low value again. The process of an abrupt rise followed by a gentle decrease repeats again and again but without obvious characteristic length as time goes on. The phenomenon is similar to the collapse of a sand pile with sands being added continuously from the top. We define the scale of collapse of the population variance by the difference of the variance before and after an abrupt rise. We then find that the logarithmic plot of the frequency versus scale of the collapses follows a power law.

## I. Introduction: the concept of self-organized criticality

Suppose the grains of sands are dropped from above the center of a sand pile. The pile will be build up gradually with a conical surface. When the slope of the surface reaches a certain maximal slope, adding sands further will induce avalanches of various sizes. Bak et al[1] simulated the process of the formation of a sand pile and found that the distribution of the size of avalanches shows a power law. The sand pile will repeatedly self-organize itself to the critical state with maximal slope and then collapse in various scales. Many other systems such as earthquakes[2], granular stick-slip system[3], traffic jam[4] were also shown to have self-organization behavior. Even evolution of life or economic growth is suspected to be a self-organized critical system[5].

The distinct features of the self-organized critical systems are the existence of a critical state to which the system will self-organize itself, and a power law distribution of certain physical quantity. In this article, we present our finding that the minority game (MG)[6] will behave as a self-organized critical system if the constraint of limiting number of agents in changing strategy is imposed. In particular, we show that the constrained MG self-organizes itself to a critical state, and the critical state collapses irregularly in various scales. The distribution of the collapse scale shows a power-law relation.

In Sec. II we introduce the standard MG and describe some of its properties that lead to our discovery of its self-organization behavior described in Sec. III. In Sec. IV we make a link between the constrained MG and a sand pile model to show they behave similarly so that the self-organized criticality appears in one system if it does in the other. Sec. V is the conclusion.

## II. The standard Minority Game

The MG was recently introduced to simulate the adaptive evolution of a population of agents. The game is played by $N$ agents in choosing one of two decisions. Those who are in the minority are rewarded one point. When the game is played repeatedly, agents use strategies in order to gain more points. The standard MG defines a strategy as a list of $2^M$ entries, each prescribes in what decision to make according to the winning history of last *M* time steps. The winning history is updated according to current outcome on every time step. A basic property of the game is the variation of population in the minority group as time goes on. The standard deviation $\sigma$ to the mean value *N*/2 is a convenient measure of how effective the system is at distributing resources. The smaller $\sigma$ is, the larger the global gain is. A remarkable result[7] was found that variance per agent, $\sigma^2/N$, is a function of $\rho = 2^M/N$.

Suppose *N* agents are assigned *S* strategies each arbitrarily in the beginning of the game. Each agent *i* uses one of his strategies $s_i$ to play the game at each time step. There exists a best set { $s_i$, *i* =1,2, …, N} such that average of the variance over all possible histories (i.e., all entries of the strategies) is minimal. We determine this minimal value $\sigma^2_{min}$ by picking a best strategy (which makes average variance smallest) for every agent once at a time and stop the process when a stable minimum is found. The results of $\sigma^2_{min}$ are plotted as a function of $\rho$ in Fig. 1.

In real game, each agent has neither enough information nor intention to choose a strategy that makes average variance minimal. Instead, he will choose the strategy which seems the best to him. A possible way which allows each agent to judge which strategy is best to him is that each agent assigns a virtual point to every strategy he has which makes correct prediction. The best strategy to an agent is then simply the one with highest virtual score. Now each agent chooses his highest score strategy each time to play the game, what can one expect for average variance? Savit et al[7] found out that average variance is slightly better (i.e. smaller) than that of not using strategy at all when $\rho$ is large, but is much worse (i.e. larger) when $\rho$ is small(Fig. 1). This comes as something of a surprise. Presumably, when each agent uses his highest score strategy, he will win the game more often than not using strategy. If all agents gain more, the global gain increases, so average variance should decrease. The presumption that one switches to a higher score strategy will win the game more often in the subsequent time steps actually fails because other agents might switch strategies in the same time. In simulation we found that many agents switch strategies simultaneously when $\rho$ is small[8], so they are likely to form majority and therefore lose the game.

To test the effect of the number of agents who switch strategies, we play the game by limiting one agent to switch strategy at each time step and calculate average variance per agent. We indeed see average variance decreases to a low value close to its minimum $\sigma^2_{min}$ for any $\rho$ (Fig. 1).

### III. The state with minimal variance is critical

Playing the MG by allowing only one agent to switch strategy at a time, average variance will gradually reach a low value close to $\sigma^2_{min}$. Keep playing the game in the same way, average variance stays near the minimal value with small fluctuation for some time. But then it rises sharply now and then with no characteristic size[Fig. 2]. The long term variation of average variance behaves very like the avalanches of a sand pile. The minimal variance state is the critical state which the system will self-organize itself to approach. But it is unstable. A small random perturbation caused by an agent's move of changing his strategy might lead other agents to change strategies in the direction to the verge of collapse, which is, in current case, a large amount of increase in variance. After a sudden collapse of certain size, the system approaches gradually the critical state again. The process repeats again and again. We define the size of collapse by the difference of the variance right before an abrupt rise and the peak value of the rise. A logarithmic plot of the frequency of occurrence versus size of the collapses for the case of $M = 6$ is given in Fig. 3(a). We see that the plot obeys a power law up to a certain size of collapse: $f(x) \propto x^{-\alpha}$, $\alpha \approx 1$ with an exponential cutoff. When the size of the system, that is *N* in the game, is larger, the power law is valid for a larger range. This is consistent with experimental data for the avalanches of a pile of rice[9], and also with the analysis of avalanches models[10]. The value of $\alpha$ is function of *M* only, independent of the size of the system (*N*). A similar plot for the case *M* = 5 is given in Fig. 3(b). Its $\alpha$ value is about 3.5.

## IV. Similarity between the sand pile and the constrained minority game

The power-law behavior of the constrained minority game which allows only one agent to change strategy at a time comes naturally when we make a similarity connection between the game and a sand pile model. Consider a sand pile built up by adding a grain of sand particle per unit time from the top. The newly added grain of sand rolls down the hill of the sand pile seeking a stable place to stay. This process is similar to the constrained MG in searching an agent to change strategy. Once the sand particle stops at some point, it changes the stability condition of at least its neighborhood. Similarly in the MG, when an agent changes his strategy, his predictions at some other histories change accordingly so that it will affect the searching results at subsequent time steps. Consider the sand pile is confined in a cylindrical container, the radius of lower surface defines the size of the sand pile inside. A grain of sand which falls from the top goes in random directions to the bottom. The size and direction play the similar roles of *N* and *M* respectively in the MG.

Keeping adding sand particles, the sand pile will reach a state with a maximal slope. Adding sands further will the induce avalanches of all sizes. In MG, changing strategy one at a time, the system will reach a state with minimal variance in population. Further changes of strategy will induce sudden increases in variance of various sizes. Both systems have a power-law behavior.

To strengthen the similarity between the sand pile system and the constrained MG, let us consider the effect of the searching process in both cases on the critical value. In the case of sand pile confined in the container, we drop sands randomly from the top within central area of radius *r*. We would expect the critical value, which is the maximal slope of the surface in this case, will be a monotonic decreasing function of *r*, since the slope has to be zero when *r* is equal to the radius of the cylinder. Our simulation using a sand-pile model similar to the one used by Bak et al[1] confirms this expectation(Fig. 4(a)). For the MG case, we first order all the agents. At each time step, we begin randomly at the *r*th agent, $r \leq N$, and search downward for one agent to change strategy. The variance will approach a critical value which is a minimum repeatedly. The minimal variance per agent is found to be monotonic increasing function of *r*(Fig. 4(b)). This is consistent with the picture that the state with the minimal variance in the MG is corresponding to the state with the maximal slope in the sand pile.

We are currently investigating the dependence of the probability density on the parameters *N*, *M*, and its possible relation to real data of earthquakes, avalanches of sand piles[9], rice piles[10], and granular stick-slip experiment[3].

## V.  Conclusion

The minority game was originally invented to simulate the variation of a system in which individual competes with one another and adapts to the global change. In the standard game, all agents change his strategy to the best one in hand at each time. We add a constraint by allowing only one agent to change strategy each time and find that the constrained minority game behaves like a self-organized critical system. In particular, we see that the variance of the number of the winning agents is attracted to a minimal value, and it arises intermittently to some large values. A power law is observed in the distribution of the variance. A comparison of the constrained minority game with the building process of a sand pile, which is generally accepted to be a typical self-organized system, shows that they both behave qualitatively the same. There are only three free parameters (*N*, *M*, and *S*) in the constrained minority game, and its statistical behavior can be understood based on the probability theory. Thus it seems possible to develop a calculable model of the self-organized critical systems using the constrained minority game.

**Acknowledgements**

This work is supported by grants from National Science Council under the grant number NSC93-2112-M005-002.

**Figure captions**

Fig. 1

Variance per agent $\sigma^2/N$ as function of $\rho = 2^M/N$ in the minority game. circle: values for standard game; cross: allow only one agent to change strategy at a time; squares: minimal values for random initial strategies. dotted line: no strategies are used. ($S=2$)

Fig. 2

A typical time sequence of the population variance as a function of time when the constraint of allowing only one agent to change his strategy is imposed. In the beginning short time period, the variance decreases to a minimal value. It then rises to a higher value intermittently ($N=1025, M=6, S=2$).

Fig. 3

Occurrence probability as a function of the size of sharp change in variance shows a power law to a certain range. The slope of the log-log plot is independent on the values of N. (a) $M=6$ for $N=2049$ (circle), $N=1025$ (cross), and $N=513$ (dot); (b) $M=5$ for $N=1025$ (circle), $N=513$ (cross), and $N=257$ (dot).

Fig. 4

Critical value as a function of the size $r$ of initial randomness. (a) For a sand pile confined in a box of size $N=50$, the maximal slope of sand pile surface is a monotonic increasing function of radius $r$ of central dropping area. (b) For the constrained MG ($N=1025$, $M=5$, $S=2$), the minimal value of variance is a monotonic increasing function of the initial searching range $r$

Fig. 1

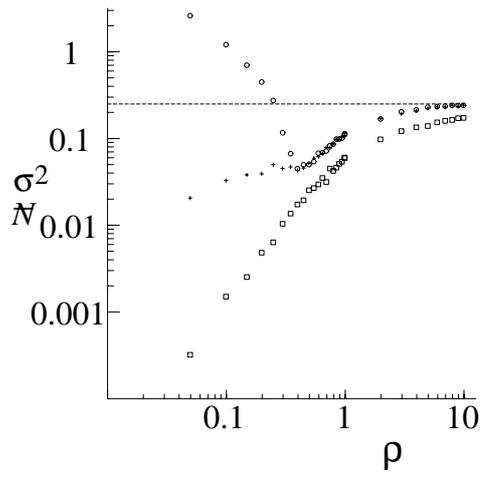

Fig. 2

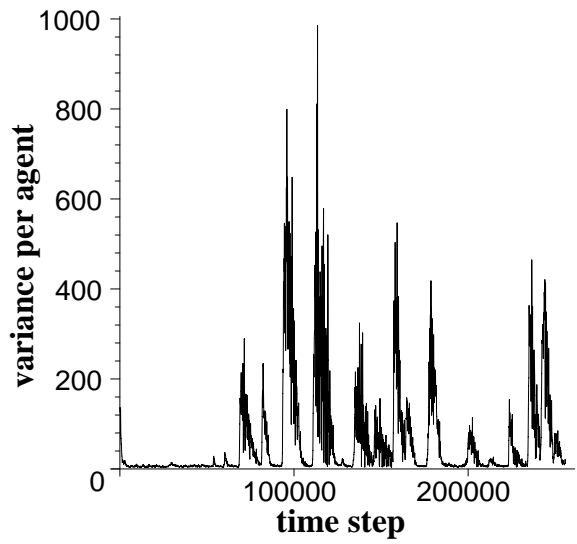

Fig. 3a

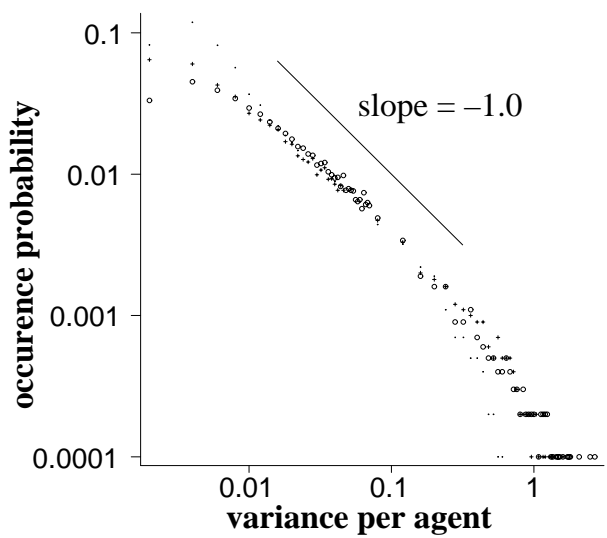

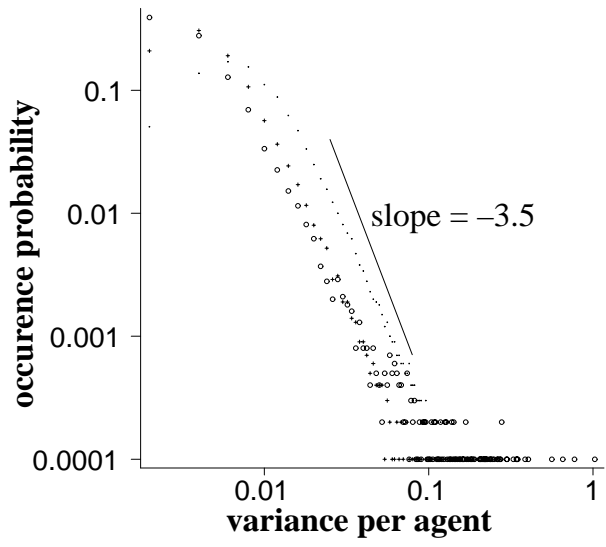

Fig. 3b

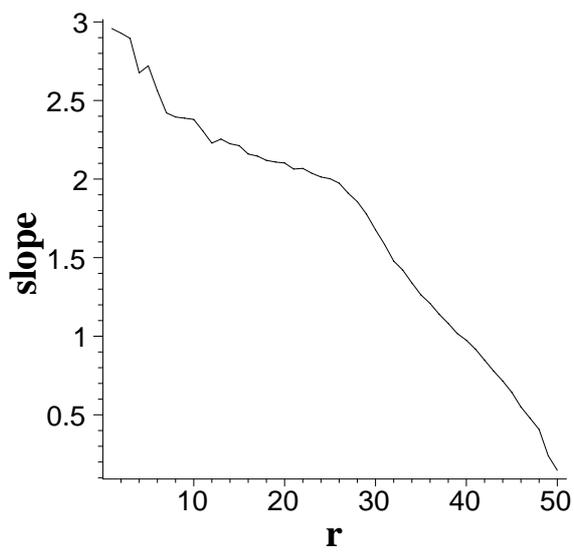

Fig. 4a

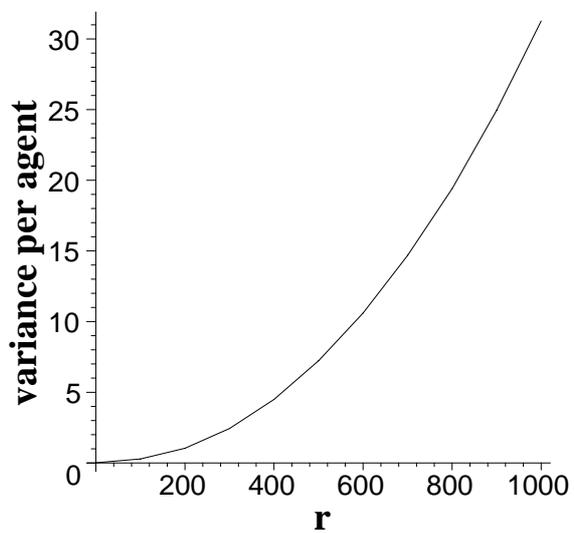

Fig. 4b